\documentclass[12pt,leqno]{amsart}
\usepackage{amssymb}
\usepackage{amsmath,amssymb}
\begin{document}
\theoremstyle{plain}
\newtheorem{MainThm}{Theorem}
\newtheorem{thm}{Theorem}[section]
\newtheorem{clry}[thm]{Corollary}
\newtheorem{prop}[thm]{Proposition}
\newtheorem{lem}[thm]{Lemma}
\newtheorem{deft}[thm]{Definition}
\newtheorem{hyp}{Assumption}
\newtheorem*{ThmLeU}{Theorem (J.~Lee, G.~Uhlmann)}

\theoremstyle{definition}
\newtheorem{rem}[thm]{Remark}
\newtheorem*{acknow}{Acknowledgments}
\numberwithin{equation}{section}
\newcommand{\eps}{{\varphi}repsilon}
\renewcommand{\d}{\partial}
\newcommand{\re}{\mathop{\rm Re} }
\newcommand{\im}{\mathop{\rm Im}}
\newcommand{\R}{\mathbf{R}}
\newcommand{\C}{\mathbf{C}}
\newcommand{\N}{\mathbf{N}}
\newcommand{\D}{C^{\infty}_0}
\renewcommand{\O}{\mathcal{O}}
\newcommand{\dbar}{\overline{\d}}
\newcommand{\supp}{\mathop{\rm supp}}
\newcommand{\abs}[1]{\lvert #1 \rvert}
\newcommand{\csubset}{\Subset}
\newcommand{\detg}{\lvert g \rvert}
\newcommand{\ppp}{\partial}
\newcommand{\dd}{\mbox{div}\thinspace}
\newcommand{\uu}{{\bf u}}

\title
[two inverse boundary value problems
for the Navier-Stokes equations]
{Equivalence of two inverse boundary value problems
for the Navier-Stokes equations}

\author{
O.~Yu.~Imanuvilov and \,
M.~Yamamoto }
\thanks{ Department of Mathematics, Colorado State
University, 101 Weber Building, Fort Collins, CO 80523-1874, U.S.A.
e-mail: {\tt oleg@math.colostate.edu}\
Partially supported by NSF grant DMS 1312900}\,

\thanks{ Department of Mathematical Sciences, The University
of Tokyo, Komaba, Meguro, Tokyo 153, Japan e-mail:
{\tt myama@ms.u-tokyo.ac.jp}}

\date{}
\maketitle
\begin{abstract}
In this note, we prove that for the Navier-Stokes equations,
a pair of Dirichlet and Neumann data and pressure uniquely correspond to
a pair of Dirichlet data and surface stress on the boundary.  Hence the two inverse
boundary value problems in [2] and [3] are the same.
\end{abstract}

\baselineskip 18pt

In Imanuvilov and Yamamoto [2],
we prove the global uniqueness in determining the viscosity
$\mu(x)$ by all Cauchy data for the Navier-Stokes equations in
two dimensions.
More precisely we can state the result as follows.
Let $\Omega \subset \Bbb R^2$ be a bounded
domain with smooth boundary $\partial\Omega$ and $\nu=(\nu_1,\nu_2)$ be
the outward unit normal vector to $\partial\Omega$ and
$\ppp_{\nu}v = \nabla v\cdot \nu$.
We set $x = (x_1, x_2) \in \Bbb R^2$ and $\Bbb N_+ = \{0,1,2,3,....\}$,
$\beta = (\beta_1, \beta_2) \in (\Bbb {N}_+)^2$, and
$\vert \beta\vert = \beta_1 + \beta_2$,
$\partial_x^{\beta} = \ppp_1^{\beta_1}\ppp_2^{\beta_2}$,
$\ppp_i = \frac{\partial}{\partial x_i}$, $i=1,2$.
We consider
the stationary Navier-Stokes equations:
\begin{align*}
&G_\mu(x,D)({\bf u},p) := \Biggl(\sum_{j=1}^2
(-2\ppp_j(\mu(x)\epsilon_{1j}({\bf u})) + u_j\ppp_ju_1)
+ \ppp_1p,\\
&\sum_{j=1}^2 (-2\ppp_j(\mu(x)\epsilon_{2j}({\bf u}))
+ u_j\ppp_ju_2 ) + \ppp_2p \Biggr) =0\quad\mbox{in}
\,\,\Omega,
\end{align*}
where ${\bf u}=(u_1,u_2)$ is a velocity field, $p$ is a pressure
and $\epsilon_{ij}({\bf u})
=\frac 12 (\ppp_iu_j + \ppp_ju_i)$.  Let
$\epsilon({\bf u}) = ( \epsilon_{ij}({\bf u}))_{1\le i,j\le 2}$.

We define the Dirichlet-to-Neumann map $\Lambda_\mu$:
$$
\Lambda_\mu({\bf f}) = (\ppp_{\nu}{\bf u}, p)\vert_{\partial\Omega},
$$
where $G_\mu(x,D)({\bf u},p)=0$ in $\Omega$, ${\bf u} = {\bf f}$
on $\partial\Omega$,
$\mbox{div}\,{\bf u}=0$, ${\bf u}\in W_2^2(\Omega)$, $p \in W_2^1(\Omega)$,
and
$$
D(\Lambda_\mu)\subset\mathcal X= \left\{{\bf f}\in
W_2^\frac 32(\partial\Omega);\thinspace \exists
 (\mbox{\bf w},q)\in W^2_2(\Omega)\times W^1_2(\Omega),
\Delta \mbox{\bf w}+\nabla q=0, \quad \mbox{div}\,\mbox{\bf w}=0,\quad
\mbox{\bf w}\vert_{\partial\Omega}=\mbox{\bf f}\right\}.
$$
The global uniqueness was proved in [2]:
\\
{\bf Theorem}
{\it We assume that $\mu_1,\mu_2\in C^{10}(\overline\Omega)$, $> 0$
on $\overline{\Omega}$ and
$\partial_x^{\beta}\mu_1 = \partial_x^{\beta}\mu_2$
on $\partial\Omega$ for each multi-index $\beta$ with
$\vert \beta\vert \le 10$.
If there exists a positive constant $\delta$ such that $$\Lambda_{\mu_1}
(\mbox{\bf f}) = \Lambda_{\mu_2}(\mbox{\bf f})\quad \forall \mbox{\bf f}\in
\mathcal X\cap \{\Vert \mbox{\bf f}\Vert_{W^\frac 32_2(\partial\Omega)}
\le \delta\},
$$
then
$\mu_1=\mu_2$ in $\Omega$.
}
\\

After [2], the paper Lai, Uhlmann and Wang [3]
appeared and proved the global uniqueness
for an inverse boundary value problem for the same Navier-Stokes
equations by using Cauchy data
$$
( {\bf u}, \sigma({\bf u},p)\nu)\vert_{\ppp\Omega},
$$
where $E_2$ is the $2\times 2$ identity matrix and
the stress tensor $\sigma({\bf u}, p)$ is defined by
$\sigma({\bf u}, p) = 2\mu(x)\epsilon({\bf u}) - pE_2$.
Our Cauchy data requires the information of the pressure $p$ on
$\ppp\Omega$, and in [3], it is written that the measurement
of $p$ alone on $\ppp\Omega$ is unnatural.

However the Cauchy data in [3] are equivalent to our data
in [2].  More precisely\\

{\bf Lemma}.
{\it Let a subbounfary $\Gamma$ of $\ppp\Omega$ be described
by $\{ (x_1, \gamma(x_1));\thinspace x_1 \in I\}$ with some
open interval $I$ and $\gamma \in C^2(\overline{I})$.
Then there exists an invertible $4 \times 4$ matrix
$K(x)\in C^1(\overline{I})$ such that
$$
K(x_1)\left(
\begin{array}{rl}
&\ppp_1({\bf u}\vert_{\Gamma})\\
& \sigma({\bf u},p)\nu\\
\end{array}
\right)
= \left(
\begin{array}{rl}
&\ppp_1u_1\vert_{\Gamma}\\
&\ppp_1u_2\vert_{\Gamma}\\
&\ppp_2u_1\vert_{\Gamma}\\
&p\vert_{\Gamma}
\end{array}
\right), \quad x_1 \in I,
$$
where ${\bf u} \in C^1(\overline{\Omega})$ satisfies
div ${\bf u} = 0$ on $\ppp\Omega$ and $p \in C(\overline{\Omega})$.
}
\\

Dividing $\ppp\Omega$ into several small subboundaries, in view of
div ${\bf u} = 0$ on $\ppp\Omega$, we see by the lemma that
$({\bf u}, \sigma({\bf u},p)\nu)\vert_{\ppp\Omega}$
uniquely determines
$({\bf u}, \ppp_{\nu}{\bf u}, p)\vert_{\ppp\Omega}$, so that
the inverse boundary value problem in [3] is the same as [2].
The same relation holds for the three dimensional case.

We note that also for the isotropic Lam\'e system,
a pair of surface displacement and Neumann derivative uniquely
corresponds to a pair of surface displacement and surface stress,
which can be proved similarly to Lemma 6.1 in Ikehata, Nakamura and
Yamamoto [1].
\\
\vspace{0.2cm}
\\
{\bf Proof.}
Without loss of generality, we can assume that $\Omega$ is located locally
below $x_2 = \gamma(x_1)$ and so we have
$$
\nu(x_1) = \frac{1}{\theta(x_1)}
\left(
\begin{array}{rl}
&-\gamma'(x_1)\\
& 1\\
\end{array}
\right),
$$
where we set $\gamma'(x_1) = \frac{d\gamma}{dx_1}(x_1)$
and $\theta(x_1) = \sqrt{1+\gamma'(x_1)^2}$.
%
Here by the divergence free condition we have
$$
(\ppp_2u_2)(x_1,\gamma(x_1)) = -(\ppp_1u_1)(x_1,\gamma(x_1)),
\quad x_1 \in I.
$$
Set $g(x_1) = u_1(x_1,\gamma(x_1))$ and $h(x_1) = u_2(x_1,\gamma(x_1))$.
Therefore
$$
(\ppp_1u_1)(x_1,\gamma(x_1)) + \gamma'(x_1)(\ppp_2u_1)(x_1,\gamma(x_1))
= g'(x_1)
                                                     \eqno{(1)}
$$
and
$$
(\ppp_1u_2)(x_1,\gamma(x_1)) - \gamma'(x)(\ppp_1u_1)(x_1,\gamma(x_1))
= h'(x_1).
                                                     \eqno{(2)}
$$
On the other hand, by the definition, we have
$$
\sigma(\uu,p)\nu = \theta(x_1)^{-1}
\left(
\begin{array}{rl}
&-2\gamma'\mu(\ppp_1u_1)(x_1,\gamma(x_1)) + \mu(\ppp_1u_2) + \mu(\ppp_2u_1)
+ p\gamma'\\
& -2\mu(\ppp_1u_1)(x_1,\gamma(x_1)) - \gamma'\mu(\ppp_1u_2)
- \gamma'\mu(\ppp_2u_1) - p \\
\end{array}\right).
$$
Setting $\left(\begin{array}{rl}
&q_1(x_1)\\
&q_2(x_1)\\
\end{array}\right)
= \theta(x_1)\sigma(\uu,p)\nu$ for $x_1 \in I$,
we have
$$
-2\gamma'\mu(\ppp_1u_1)(x_1,\gamma(x_1)) + \mu(\ppp_1u_2) + \mu(\ppp_2u_1)
+ p\gamma'(x_1,\gamma(x_1)) = q_1(x_1)                \eqno{(3)}
$$
and
$$\\
-2\mu(\ppp_1u_1)(x_1,\gamma(x_1)) - \mu\gamma'(\ppp_1u_2)
- \mu\gamma'(\ppp_2u_1) - p(x_1,\gamma(x_1)) = q_2(x_1),       \eqno{(4)}
$$
Setting $f_1(x_1) = (\ppp_1u_1)(x_1,\gamma(x_1))$,
$f_2(x_1) = (\ppp_1u_2)(x_1,\gamma(x_1))$, $f_3(x_1)
= (\ppp_2u_1)(x_1,\gamma(x_1))$
and $f_4(x_1) = p(x_1)$, we can rewrite (1) - (4) as
$$
A(x_1)\left(
\begin{array}{rl}
&f_1\\
&f_2\\
&f_3\\
&f_4\\
\end{array}
\right)
:=
\left(
\begin{array}{cccc}
1        & 0     & \gamma' &0 \\
-\gamma' & 1     & 0       &0 \\
-2\mu\gamma' &\mu & \mu & \gamma' \\
-2\mu& -\mu\gamma' & -\mu\gamma' &-1
\end{array}
\right)
\left(
\begin{array}{rl}
&f_1\\
&f_2\\
&f_3\\
&f_4\\
\end{array}
\right)
= \left(
\begin{array}{rl}
&g'\\
&h'\\
&q_1\\
&q_2\\
\end{array}
\right),
\quad x_1 \in I.
$$
Multiplying the fourth row of $A$ with $\gamma'$ and adding to the
third row, we have
$$
\mbox{det}\thinspace A
= -\mbox{det}\thinspace
\left(
\begin{array}{cccc}
1        & 0     & \gamma' \\
-\gamma' & 1     & 0  \\
-4\mu\gamma' &\mu(1-\vert\gamma'\vert^2) & \mu(1-\vert\gamma'\vert^2) \\
\end{array}
\right)
= (\vert \gamma'\vert^4 + 2\vert \gamma'\vert^2 +1)\mu
\ne 0.
$$
Since $\theta(x_1) = \sqrt{1+\gamma'(x_1)^2} > 0$ for $x_1 \in
\overline{I}$, the invertibility of $A$ completes the proof of
the lemma.

\end{document}